\documentclass[a4paper,aps,prd,showpacs,preprintnumbers,twocolumn,nofootinbib,superscriptaddress]{revtex4-1}

\usepackage[normalem]{ulem} 
\usepackage{graphicx}
\usepackage{amsmath}
\usepackage{amsfonts}   
\usepackage{amssymb}  
\usepackage{mathrsfs}
\usepackage{bm}
\usepackage{color}

\newcommand{\dd}{{\rm d}}
\newcommand{\mpg}{M_g}
\newcommand{\mpf}{M_f}
\newcommand{\binomial}[2]{ \left(\begin{array}{c} #1 \\ #2 \end{array}\right)}

\begin{document}

\title{Extending applicability of bimetric theory: chameleon bigravity}

\author{Antonio De Felice}
\email{antonio.defelice@yukawa.kyoto-u.ac.jp}
 \affiliation{
             Center for Gravitational Physics, Yukawa Institute for Theoretical Physics, Kyoto University, 606-8502, Kyoto, Japan}

\author{Shinji Mukohyama}
\email{shinji.mukohyama@yukawa.kyoto-u.ac.jp}
 \affiliation{
             Center for Gravitational Physics, Yukawa Institute for Theoretical Physics, Kyoto University, 606-8502, Kyoto, Japan}
\affiliation{
              Kavli Institute for the Physics and Mathematics of the Universe (WPI), The University of Tokyo Institutes for Advanced Study, The University of Tokyo, Kashiwa, Chiba 277-8583, Japan}

\author{Jean-Philippe Uzan}
\email{uzan@iap.fr}
 \affiliation{
            Institut d'Astrophysique de Paris, CNRS UMR 7095, Universit\'e Pierre \& Marie Curie - Paris VI, 98 bis Bd Arago, 75014 Paris, France \\
           Sorbonne Universit\'es, Institut Lagrange de Paris, 98 bis, Bd Arago, 75014 Paris, France.}

\begin{abstract}
This article extends bimetric formulations of massive gravity to make the mass of the graviton to depend on its environment. This minimal extension offers a novel way to reconcile massive gravity with local tests of general relativity without invoking the Vainshtein mechanism. On cosmological scales, it is argued that the model is stable and that it circumvents the Higuchi bound, hence relaxing the constraints on the parameter space. Moreover, with this extension the strong coupling scale is also environmentally dependent in such a way that it is kept sufficiently higher than the expansion rate all the way up to the very early universe, while the present graviton mass is low enough to be phenomenologically interesting. In this sense the extended bigravity theory serves as a partial UV completion of the standard bigravity theory. This extension is very generic and robust and a simple specific example is described. 
\end{abstract}
\date{\today}
\preprint{YITP17-15, IPMU17-0034}
\maketitle

\noindent{\bf\em Introduction.} Bimetric theories of gravity attract a lot of interest, in particular with the goal of designing a theoretically consistent and observationally viable theory of massive gravitons; see e.g.\ Refs~\cite{review} for reviews. Those theories consider two metrics, $g_{\mu\nu}$ and $f_{\mu\nu}$, interacting through a potential that depends on $s^\alpha_\beta\equiv (\sqrt{g^{-1}f})^{\alpha}_{\beta}$ (the square-root being defined such that $s^\mu_\alpha s^\alpha_\nu =g^{\mu\alpha} f_{\alpha\nu}$).

Nonetheless, the bimetric theory introduced in Ref.~\cite{Hassan} has three shortcomings. {\em First} of all, in order to be a dark energy candidate, one requires, in general, that the mass parameter $m$ (of order of the mass of the massive graviton) to be of order of today's Hubble parameter. However, as shown in Ref.~\cite{comelli} (see also \cite{referee1} for other problems), it implies that at early times, e.g.\ during radiation domination, the scalar perturbations would develop a short-time-scale gradient instability, which would make the cosmology unviable. Therefore one had to abandon the idea that the massive graviton would be the source of dark energy. On pursuing this latter choice, it was shown~\cite{DeFelice:2014nja} (see also \cite{Akrami:2015qga}) that it was possible to achieve a stable evolution on a cosmological background as long as $\rho_{\rm matter}/(m^2\mpg^2)\ll1$ (during both radiation and matter eras), which implies $m$ to be, in general, much larger than $H_0$. On the other hand, this same assumption must break down at some early times, so that the theory requires a UV-completion or a partial UV-completion.

Let us turn to the {\em second} problem of this theory. As long as we consider it at low energies, this theory offers a mathematically and phenomenologically consistent framework, only if there exists a mechanism to  suppress the propagation of the extra bigravity modes in the Solar system, so that local gravity constraints can be satisfied. This is usually achieved by relying on the Vainshtein mechanism~\cite{vmeca}, i.e.\ invoking the non-linear structure of the theory. On the other hand, since $m^2\gg H_0^2$, one typically expects the massive graviton (together with a massless one) to be so heavy that it can be integrated out from the theory (making it indistinguishable from GR). This would make the theory stable, but somehow rather uninteresting. Therefore, in order to have a non-trivial phenomenology, it was proposed in Ref.~\cite{adeftan} that a fine-tuning was required in order to set the effective mass of the graviton (which is proportional but not equal to $m$) low enough in order to (1) pass Solar system constraints, and (2) give a non-trivial phenomenology in terms of gravitation-wave oscillations. This fine-tuning is then introduced in addition to the fine-tuning of the cosmological constant, leading to a stable but contrived construction of a ghost-free bigravity theory.

The {\em third} problem is that the strong coupling scale of the theory, $\sim (M_g m^2)^{1/3}$, is rather low for phenomenologically interesting values of $m$. For this reason, once $m$ is chosen low enough to be phenomenologically interesting, it is difficult to apply the theory to early universe cosmology. Some sort of (partial) UV completion is necessary. 

To summarize, bigravity theories of gravity suffer from 3 shortcomings. While two of them are related to the viability of the theory (associated to the Vainshtein mechanism and Higuchi bound), the third is somehow less important and is related to the will to get a new phenomenology for this theory, which is a hope but not a necessity.

This letter  proposes a minimal extension of these theories that allows for the mass of the massive graviton to be environmentally dependent. This sole condition, we argue, will fix all three problems of the standard bigravity theories. We shall argue that this can be implemented by extending the standard chameleon mechanism~\cite{chameleon} and that it frees the model from the Vainshtein mechanism and the Higuchi bound, hence significantly broadening the regime of applicability of the bimetric theories. Moreover, with this extension the strong coupling scale is also environmentally dependent in such a way that it is kept sufficiently higher than the expansion rate all the way up to the very early universe, while the present graviton mass is low enough to be phenomenologically interesting. In this sense the extended bigravity theory proposed in the present article serves as a partial UV completion of the standard bigravity theory. 

\vskip.25cm
\noindent{\bf\em Standard bimetric theories in a nutshell.}  Bimetric theories considered so far, rely on the action
\begin{eqnarray}\label{e:action_bigrav}
 S&=&\frac{\mpg^2}{2}\int R[g]\sqrt{-g}\dd^4x + \frac{\mpf^2}{2}\int R[f]\sqrt{-f}\dd^4x\nonumber\\
       && + \mpg^2m^2 \int \sum_{i=0}^4 \beta_i U_i[s]\sqrt{-g}\dd^4x\,, 
\end{eqnarray}
where $\mpg$ and $\mpf$ are the Planck masses in the gravitational $g$ and $f$ sectors, respectively, and we define $\kappa\equiv\mpf^2/\mpg^2$. The potentials $U_i$ ($i=1,\cdots,4$) are defined in terms of $T_n\equiv {\rm Tr}[s^n]$ as $U_0=1$, $U_1=T_1$, $U_2=\frac{1}{2}\left[T_1^2- T_2\right]$, $U_3=\frac{1}{6}\left[ T_1^3 -  3T_2T_1 + 2T_3\right]$, and $U_4=\frac{1}{24}\left[ T_1^4-6T_1^2T_2+3T_2^2+8T_1T_3-6T_4\right]$, where, as already stated above, $s^\alpha_\beta\equiv (\sqrt{g^{-1}f})^{\alpha}_{\beta}$.
The cosmological constants are not included explicitly since they are already contained in $U_0$ and $U_4$. 

The cosmological analysis of this theory~\cite{DeFelice:2014nja} assumes the background to be of the form 
\begin{eqnarray}
\dd s_g^2&=&-\dd t^2+a^2(t)\delta_{ij}\,\dd x^i\,\dd x^j\,,\nonumber\\
\dd s_f^2&=&\xi^2(t)\,\left[-c^2(t)\dd t^2+a^2(t)\delta_{ij}\,\dd x^i\,\dd x^j\right]\,. \label{e.f}
\end{eqnarray}
Among the various existing branches, Ref.~\cite{DeFelice:2014nja} concluded that only the one  with $H=\xi\hat H$ (with $H\equiv \dot a/a$ and $\hat H\equiv (a\xi)^./(ac\xi^2)$) is healthy (see also \cite{comelli}). For a matter field with energy density $\rho$, minimally coupled with $g_{\mu\nu}$, the Friedmann equation takes the form
\begin{equation}
3 \mpg^2 H^2 =\rho+m^2\mpg^2 R\,, \quad R \equiv U-\xi U_{,\xi}/4\,,
\end{equation}
where $U=-(\beta_4\xi^4+4\beta_3\xi^3+6\beta_2\xi^2+4\beta_1\xi+\beta_0)$. Provided that the condition~\cite{adeftan}
\begin{equation}
 \rho\ll \mpg^2m^2\, \label{eqn:mbig}
\end{equation}
holds, there exists a stable cosmological background whose Friedmann equation behaves as in the standard cosmology, $H^2\propto\rho$ up to corrections that are negligible below a certain high energy scale. The cosmological evolution then deviates from the standard one at high energy, $\rho\gtrsim \mpg^2 m^2$.

It was also shown~\cite{DeFelice:2014nja} that the massive tensor mode has a mass
\begin{equation}\label{e.meff}
 m_{\mathrm{T}}^2 =\frac{1+\kappa\xi^2}{\kappa\xi^2}m^2\Gamma\,, \quad
  \Gamma \equiv \xi J + (c-1)\xi^2J_{,\xi}/2\,,
\end{equation}
where $J=R_{,\xi}/3$. Obviously this formula applies to the mass in the Einstein frame. Given the condition (\ref{eqn:mbig}), the mass can be made light enough ($m_{\mathrm{T}}^2 \ll m^2$) so as to give a new phenomenology detectable~\cite{adeftan} by gravitational wave detectors only if the parameters of the theory are fine-tuned so that $\Gamma\approx0$. 

This latter condition is also required for the implementation of the Vainshtein mechanism (so as to suppress any fifth force arising from the extra gravity degrees of freedom). This mechanism may work only if 
\begin{equation}
\left|\frac{{\rm d}\ln\Gamma}{{\rm d}\ln\xi}\right|\gg1\,,
\end{equation}
which generically holds for $\Gamma\approx0$. To finish, the stability of de Sitter backgrounds also requires that $m_{\mathrm{T}}^2>2H^2$, referred to as the Higuchi bound~\cite{higuchi}.

\vskip.25cm
\noindent{\bf\em Extension allowing for a scaling of the interaction potential.} Suppose the coefficients of the potential depends on the environment in such a way that they scale as (we shall discuss how to implement this idea below) $\beta_i \propto \rho$,  then it is clear that $U\propto\rho$ and $\Gamma\propto\rho$ so that $m^2_{\mathrm{T}}\propto\rho$. We then have
\begin{equation}
\frac{m_{\mathrm{T}}^{\rm loc}}{m_{\mathrm{T}}^{\rm cosm}}\approx \sqrt{\frac{\rho_{\rm loc}}{\rho_{\rm cosm}}}\sim10^5\,,
\end{equation}
if we assume that the mean density on Solar system scales is given by the dark matter density $\rho_{\rm loc}\sim\rho_{\rm DM, Solar Syst.}\simeq 1.4\times10^{-19}\, \textrm{g.cm}^{-3}$ and that the cosmological density is $\rho_{\rm cosm} = 3H_0^2\Omega_{m0}/8\pi G\sim 10^{-29}\textrm{g.cm}^{-3}$.

The requirement that the massive graviton is heavy enough on Solar system scales imposes $m_{\mathrm{T}}^{\rm loc}\gg 1 {\rm AU}^{-1}\approx8\times10^{-18}$ eV.
Combined with the assumed scaling, this sets the bound at cosmological scales as $m_{\mathrm{T}}^{\rm cosm}\gg 8\times10^{-23}$ eV.
Unfortunately, this value is too large for the present model to allow for self-acceleration.
On the other hand, for the first time in the context of cosmology in bimetric theories, the present model allows for a stable cosmology all the way
from the early radiation-domination down to the present epoch and for interesting gravitational wave phenomenology.
In particular, this constraint allows cosmological values as light as $m_{\mathrm{T}}^{\rm cosm}\simeq 10^5\, {\rm pc}^{-1}=6.4\times10^{-19}$ eV which may lead to graviton oscillations \cite{adeftan}\footnote{Actually, for this value of $m_{\mathrm{T}}^{\rm cosm}$, we should expect two waves coming to us at different arrival times: one coming at the speed of light (which is consistent with multimessenger constraints), the other one should be delayed by some amount of time which can be in principle detected by using new GWs templates. For example, for a wave with 100 Hz frequency coming from a distance of 1 Mpc, we have a $\Delta t$ of order of a few hours.}, which would not be possible otherwise without fine-tuning and abandonment of early universe cosmology. With scaling of the form $m_{\mathrm{T}}\propto \sqrt{\rho}$, the bound on the cosmological value of the mass of the massive tensor mode coming from Solar system constraints is relaxed and leaves a larger space for non-trivial cosmological phenomenology.

Tabletop experiments also impose constraints on the model. Inside a source of mass $M_{\rm source}$ and size $L_{\rm source}$, we have $m_{\mathrm{T}}^{\rm source}/m_{\mathrm{T}}^{\rm cosm}\approx \sqrt{M_{\rm source}/(L_{\rm source}^3\rho_{\rm cosm})}$. We need the graviton mass $m_{\mathrm{T}}^{\rm source}$ inside the source to be heavier than $1/L_{\rm source}$ for the Yukawa potential not to seep out from the source mass. This puts the bound,
\begin{equation}
 m_{\mathrm{T}}^{\rm cosm} \gg \sqrt{\frac{L_{\rm source}\rho_{\rm cosm}}{M_{\rm source}}}\,.
\end{equation}
Among various tabletop experiments (see e.g. Ref~\cite{Murata:2014nra}), Hoskins {\em et al.}~\cite{Hoskins:1985tn} (with $L_{\rm source}\sim 5{\rm cm}$ and $M_{\rm source}\sim 43{\rm g}$) gives the strongest bound $m_{\mathrm{T}}^{\rm cosm}\gg 2\times 10^{-20}$ eV. While slightly stronger than the constraint from the Solar system scales, it still allows cosmological values as light as $m_{\mathrm{T}}^{\rm cosm}\simeq 10^5\, {\rm pc}^{-1}=6.4\times10^{-19}$ eV.

As we shall see later, such a scaling may also alleviate the constraint required for the stability of the cosmological evolution all the way up to the scale at which the effective theory describing bigravity breaks down. The constraint due to this requirement would be the most stringent in the standard bigravity. This cosmological stability condition is often called the generalized Higuchi condition~\footnote{The bound derived in \cite{Fasiello:2013woa} is for homogeneous and isotopic perturbation and thus is the condition for the avoidance of an IR ghost, which a priori is not necessarily the right condition for theoretical consistency~\cite{Gumrukcuoglu:2016jbh}. The no-ghost condition for general inhomogeneous and anisotropic perturbation in the high $k$ limit was later derived in \cite{DeFelice:2014nja}.\label{footnote:generalizedHiguchibound}},
\begin{equation}
 m_{\mathrm{T}}^2>\mathcal{O}(1)\times H^2\,, \label{eqn:generalHiguchi}
\end{equation}
where $m_{\mathrm{T}}$ and $H$ are the mass of the massive tensor mode and the Hubble expansion rate at the time of interest. If a mechanism allows for the scaling $m_{\mathrm{T}}\propto \sqrt{\rho}$, then the generalized Higuchi condition is almost independent of the scale and thus will be automatically satisfied at all scales once it is satisfied at one scale. 

\vskip.25cm
\noindent{\bf\em Implementing the mechanism.} The simplest way to get an environmental dependence of the coefficients of the potentials $U_i$ is to promote them to functions of a scalar field~\footnote{In the context of theories of single massive graviton, the idea of promoting the coefficients of the graviton interaction terms to functions of a scalar field is not new. For example, see Ref.~\cite{DAmico:2012hia,referee2}.} and to implement the chameleon mechanism~\cite{chameleon}. This is easily achieved by extending the action~(\ref{e:action_bigrav}) by first, letting $\beta_i$ depend on a scalar field $\phi$ as
\begin{equation}
 \beta_i = \beta_i(\phi)\,,
\end{equation}
then allowing for a non-minimal coupling of the matter sector
\begin{equation}
S_{\rm mat} = \int {\cal L}_{\rm mat}(\psi,\tilde g_{\mu\nu})\sqrt{-\tilde g}\,\dd^4x\,,\label{eqn:matteraction}
\end{equation}
where $\tilde g_{\mu\nu} = A^2(\phi)g_{\mu\nu}$ and $A(\phi)$ is a coupling function, assumed to be universal, and by adding a kinetic term of the scalar field
\begin{eqnarray}
S_{\rm kin} = -\frac{1}{2}\int g^{\mu\nu}\partial_\mu\phi\partial_\nu\phi \sqrt{-g}\dd^4x.
 \label{eqn:Skin}
\end{eqnarray}
This defines an extended bigravity theory in the Einstein-frame. The Jordan-frame description of the same theory can be defined by shifting to $\tilde g_{\mu\nu}$ and $\tilde f_{\mu\nu}=A^2(\phi)f_{\mu\nu}$ that let $s$ invariant. The interaction terms $\mpg^2 m^2\sum_i\beta_i(\phi)U_i$ play the role of a potential for $\phi$ as well. Finally, we define the coupling strength of the field to matter by $\alpha(\phi)\equiv\dd\ln A/\dd\phi$.  No potential was added to (\ref{eqn:Skin}) since it can be absorbed in a redefinition of $\beta_0(\phi)$.

The dynamics of such a theory is governed by two Einstein equations, a Klein-Gordon equation and a conservation equation, which take the form
\begin{eqnarray}
 G_{\mu\nu}[g] & = & \frac{1}{\mpg^2}\left[T_{\mu\nu} + \partial_\mu\phi\partial_\nu\phi-\frac{1}{2}\partial_\alpha\phi\partial_\beta\phi g^{\alpha\beta}g_{\mu\nu}\right]\nonumber\\
 & & + m^2\sum_{i=0}^3 (-1)^{i+1}\beta_i(\phi) Y^{(i)}_{\mu\nu}\,, \nonumber\\
  G_{\mu\nu}[f] & = & m^2\sum_{i=0}^3 (-1)^{i+1}\beta_{4-i}(\phi) \hat{Y}^{(i)}_{\mu\nu}\,,\nonumber\\
 \Box\phi & = & - \alpha T_{\mu\nu}g^{\mu\nu} - \mpg^2m^2 \sum_{i=1}^3\beta'_i(\phi)U_i[s]\,,\nonumber\\
  \nabla_\mu T^{\mu\nu} & = & \alpha(\phi) T^{\mu\nu}\partial_\mu\phi\,,
\end{eqnarray}
where $T^{\mu\nu}$ is the stress-energy tensor of matter fields in the Einstein frame,
\begin{eqnarray}
   Y^{(0)}_{\mu\nu} &=& g_{\mu\nu}\,,\quad 
   Y^{(1)}_{\mu\nu} = [s^\alpha_\mu -T_1\delta^\alpha_\mu]g_{\nu\alpha}\,,\nonumber\\
   Y^{(2)}_{\mu\nu} &=& g_{\mu\alpha}s^\alpha_\beta s^\beta_\nu - T_1 g_{\mu\alpha} s^\alpha_\nu +\frac{1}{2}\left[T_1^2- T_2 \right]g_{\mu\nu}\,,\nonumber\\      
   Y^{(3)}_{\mu\nu} &=&  g_{\mu\alpha}s^\alpha_\beta s^\beta_\gamma s^\gamma_\nu - T_1 g_{\mu\alpha} s^\alpha_\beta s^\beta_\nu
    +\frac{1}{2}\left[T_1^2- T_2 \right]g_{\mu\alpha}s^\alpha_\nu \nonumber\\
 & &  - \frac{1}{6}\left[ T_1^3 -  3T_2T_1 + 2T_3\right]  g_{\mu\nu}\,,
\end{eqnarray}
and $\hat{Y}^{(i)}_{\mu\nu}$ is defined in the same way as $Y^{(i)}_{\mu\nu}$ with $g_{\mu\nu}\leftrightarrow f_{\mu\nu}$. 

On {\it cosmological scales}, adopting the ansatz (\ref{e.f}), the field equations reduce to
\begin{eqnarray}
& & 3 H^2 = \frac{1}{\mpg^2}\left[\rho A^4 +\frac{1}{2}\dot\phi^2\right] + m^2R\,,\label{eqn:Friedmann-eqs1}\\
& & 3\hat H^2 =\frac{m^2}{4\kappa\xi^3}\, U_{,\xi}\,, \label{eqn:Friedmann-eqs2}\\
& & 2\dot H = -\frac{1}{\mpg^2}\left[ (\rho+P) A^4+\dot{\phi}^2 \right] + m^2\xi(c-1)J\,,\label{eqn:Friedmann-eqs3}\\
& & 2\dot{\hat{H}} = m^2\frac{1-c}{\kappa\xi^2}\, J\,, \label{eqn:Friedmann-eqs}
\end{eqnarray}
and
\begin{eqnarray}
 \ddot\phi +3 H\dot\phi & = & -\alpha A^4(\rho-3P)+ \mpg^2m^2 Q_{,\phi}\,, \label{kg1}
\end{eqnarray}
with $Q(\xi,\phi)\equiv (c-1)\,R-c\,U$.  Note that $\rho$ is the
energy density in the Jordan frame, defined by
$T_{\mu\nu}(\partial/\partial t)^{\mu}(\partial/\partial
t)^{\nu}\equiv \rho A^4$. Hence the conservation equation implies that
$\rho\propto (Aa)^{-3}$ for pressureless matter, for example. The
functions $U$, $R$ and $J$, defined as before, now depend on both
$\xi$ and $\phi$. By combining Eqs.~(\ref{eqn:Friedmann-eqs1}) and
(\ref{eqn:Friedmann-eqs3}) with Eq.~(\ref{kg1}), it is straightforward
to derive the constraint
\begin{eqnarray}
\!\! \!\! \!\! \!\! \left[\dot{\xi}+ (1-c)H\xi\right]\sum_{i=0}^2 \binomial{2}{i}\beta_{i+1}\xi^i
 =\frac{c\xi}{3} \sum_{i=0}^3 \binomial{3}{i}\dot{\beta}_{i+1}\xi^i\, .
 \label{eqn:constraint}
\end{eqnarray}

On {\it local scales}, we assume that the spacetime is close to Minkowski and solve, in the chameleon spirit~\cite{chameleon}, the Klein-Gordon equation
\begin{equation}
 \frac{\dd^2\phi}{\dd r^2}+\frac{2}{r}\frac{\dd\phi}{\dd r} = \rho\alpha A^4 - \mpg^2m^2 Q_{,\phi}\,,\label{kg2}
\end{equation}
assuming that the local density is $\rho=\rho_{\rm loc}$ and $\xi=\xi_{\rm loc}$.

\vskip.25cm
\noindent{\bf\em Mechanism.} For simplicity, and as a proof of concept, we assume that
\begin{equation}\label{e.hyp}
 \beta_i(\phi) = -c_i f(\phi)\,,
\end{equation}
so that $U[\beta_i(\phi),\xi]\rightarrow -U[c_i,\xi]f(\phi)$ since it is a linear function of the coefficients $\beta_i$. The same applies to the functions $R$, $J$ and $Q$. The universal coefficient function $f(\phi)$ is chosen as a decreasing function of $\phi$ and the coupling function is assumed to be given by
\begin{equation}
A=\hbox{e}^{\beta \phi/\mpg}\,,
\end{equation}
so that $\alpha=\beta/\mpg$ is a constant. Now, in the local region, $\xi=\xi_{\rm loc}$ is constant and $c=1$. On cosmological scales it can be shown~\cite{DeFelice:2014nja} that $\xi\rightarrow \xi_c$ constant and $c\rightarrow 1$. We can thus argue that $\xi_{\rm loc}=\xi_c$ and $c=1$ at late time and on cosmological scales (denoted by infinity). 

The potential appearing in the Klein-Gordon equations~(\ref{kg1}) and~(\ref{kg2}) has a minimum at a value $\phi=\phi_*(\rho,\xi,c)$, which will mostly be a function of $\rho$ under the above conditions. Hence, $m_{\mathrm{T}}^2\propto f(\phi)$ is essentially a function of $\rho$. This clearly realizes the idea discussed above. The exact scaling of $\phi_*$, and thus $m_{\mathrm{T}}^2$, with $\rho$ depends on the form of the function $f(\phi)$ so that we need to take more specific examples to put numbers. Nonetheless this discussion at the very least demonstrates the generality of the mechanism. It can be easily extended by allowing for each $\beta_i$ to have a different scaling.

\vskip.25cm
\noindent{\bf\em Example.}  Let us further assume the form
\begin{equation}
 f=\hbox{e}^{-\phi/M}=\hbox{e}^{-\lambda \phi/\mpg}\,, \label{eqn:1stexample}
\end{equation}
where $M$ is a mass scale and $\lambda=\mpg/M$. Clearly, the local and cosmological values of the field are related by
\begin{equation}\label{e.ff}
(\lambda + 4\beta)(\phi_\infty-\phi_{\rm loc}) = \mpg\ln\frac{\rho_{\rm loc}}{\rho_\infty}\,
\end{equation}
as long as a backreaction is small and assuming the field has settled at the minimum of its potential. We conclude that the masses of the massive tensor mode at local and cosmological scales are related by
\begin{equation}
 \frac{m^2_{\rm T,loc} }{m^2_{\rm T,\infty}}= \left(\frac{\rho_{\rm loc}}{\rho_\infty} \right)^{\frac{\lambda}{\lambda+4\beta}}\,.
\end{equation}
Assuming $\beta={\cal O}(1)$ and $M\ll \mpg$ (so that $\beta\ll \lambda$), we end up with the announced scaling $m_{\rm T}^2\propto \rho$. The situation is then similar as in the standard chameleon mechanisms~\cite{chameleon}.

\vskip.25cm
\noindent{\bf\em 5th force.} While the fifth force associated with the gravity degrees of freedom is suppressed by our mechanism, the scalar field $\phi$ is also responsible for such a force (so to speak the 6th force). Equation~(\ref{kg2}) is similar to the standard chameleon case~\cite{chameleon} but with a potential $V= -\mpg^2m^2Q(\xi,\phi)=-\mpg^2m^2Q(\xi,0)\hbox{e}^{-\lambda \phi/\mpg}$. Around a massive body of density $\rho_c$ and radius $R_c$, the field will have a profile interpolating from $\phi_c$ inside to $\phi_{\rm loc}$ outside, these two values being related by the same relation as Eq.~(\ref{e.ff}). For the mechanism to apply, we need to have $Q(\xi,0)<0$, which can be achieved if $c_i>0$ and $c\simeq 1$ for example. Asymptotically, the mass of the scalar field is given by the second derivative of the effective potential at its minimum. In the limit $\beta\ll\lambda$ it is well approximated by $m_\phi^2\sim \lambda\beta\rho/\mpg^2$. The field becomes massive in high density environment and the theory can pass tests on fifth force. The amplitude of the fifth force depends on the density and size of the objects and environment as well as the parameter of the model, and in particular if the thin shell approximation holds. Indeed, depending on the parameters, violations may be detected in space. We hope to detail the full analysis of the parameter space in a future publication. 

In (\ref{eqn:matteraction}), the scalar field $\phi$ is assumed to couple universally to matter ensuring that the weak equivalence principle holds~\cite{Uzan:2010pm}. Therefore, the model passes all 5th force experimental constraints as far as $m_{\phi}$ is heavy enough. Since we have the relation $m_{\phi}^2\propto \rho$, the strongest bound again comes from Hoskins {\em et al.}~\cite{Hoskins:1985tn}. By requiring $m_\phi^2L^2_{\rm source} \sim \lambda\beta M_{\rm source}/(\mpg^2L_{\rm source})\gg 1$ and supposing $\mpg\simeq M_{\rm Pl}$, we obtain the constraint $M\ll\beta\times 10\, {\rm eV}$. We omit here the study of other complications related to the details of the experiments, e.g.\ their geometry, boundaries, and other details regarding extended sources for the chameleon screening mechanism, as we believe these issues should be addressed in a study explicitly devoted to test the theory in an actual experiment, and as such, to be explained in a future paper.

\vskip.25cm
\noindent{\bf\em Stability on de Sitter. }  We want to know whether the final state of the universe would be stable or not. To address this issue, we assume the universe is asymptotically well-described by a de Sitter background with $\phi=const$, and thus suppose that the matter enjoys the equation of state $P=-\rho$. As in the standard bigravity, two different branches exist since for $\dot{\phi}=0$ the constraint equation (\ref{eqn:constraint}) reduces to $(1-c)(\beta_3\xi^2+2\beta_2\xi+\beta_1)=0$.  We shall focus on the normal branch, i.e.\ the one which does not suffer from strong-coupling and/or ghosts. It is defined by the condition $c=1$. For the example~(\ref{eqn:1stexample}), the Friedmann equations give
\begin{eqnarray}
H^2 &=& \frac{m^2}{3}\,e^{-\frac{\lambda\,\phi}{\mpg}}\sum_{i=0}^3\binomial{3}{i}c_i\xi^i+\frac{\rho}{3\mpg^2}\,e^{\frac{4\beta\,\phi}{\mpg}}\,,\\
H^2&=&\frac{m^2}{3\kappa\xi}\, e^{-\frac{\lambda\phi}{\mpg}} \sum_{i=0}^3\binomial{3}{i}c_{i+1}\xi^i\,,
\end{eqnarray}
where both $\xi$ and $\phi$ have constant values. After using Eq.~(\ref{kg1}) to eliminate $\rho$, it implies
\begin{eqnarray}
\beta\sum_{i=0}^3\binomial{3}{i}\xi^i\left(c_i\xi-\frac{c_{i+1}}{\kappa}\right) 
+ \frac{\lambda\xi}{4}\, \sum_{i=0}^4\binomial{4}{i}c_i\xi^i  = 0\,.\nonumber\\
\label{eq:fxi}
\end{eqnarray}
This algebraic constraint can be solved to get $\xi$ as a function of ($c_i$, $\lambda$, $\beta$, $\kappa$). The solutions depend neither on $\phi$ nor on $\rho$, but only on the parameters of the theory. In other words, they do not depend on the environment.

Let us now study the stability of this background against cosmological perturbations. Both tensor modes (with two polarizations each) propagate with the speed of light (for high $k$'s). However, one can show that the eigen-mass spectrum contains one eigen-state with a vanishing squared-mass eigenvalue  (i.e.\ one of the gravitons remains massless), whereas the other one satisfies 
\begin{equation}
\frac{m_{\mathrm{T}}^2}{H^2} = \frac{3(1+\kappa\,{\xi}^{2})(c_3\xi^2+2c_2\xi+c_1)}{c_4\xi^3+3c_3\xi^2+ 3c_2\xi+c_1}\,. \label{eqn:mT2}
\end{equation}
This explicitly shows that $m_{\mathrm{T}}^2/H^2$, being only a function of $\xi$, is independent of the environment. This leads to the crucial fact that, the larger $H$ the larger the mass of the graviton. We also found that the no-ghost condition for the vector modes requires that $m_{\mathrm{T}}^2>0$, whereas their speed of propagation (for high $k$'s) turns out to be unity (as expected from this de Sitter-invariant background). Finally the scalar modes can be shown to have two modes propagating with speed of light (for high $k$'s), which are both not ghosts if the following relation holds
\begin{equation}
 \frac{m_{\mathrm{T}}^2}{H^2} > 1 + \sqrt{1+6\lambda^2\kappa\xi^2(1+\kappa\xi^2)}\,.
  \label{eqn:noghost-scalar}
\end{equation}
For $\lambda\ne 0$ (experimental constraints indeed require $\lambda\gg\beta$ but $\kappa$ may be small), this is a relation stronger than the usual Higuchi bound,
\begin{equation}
 m_{\mathrm{T}}^2>2H^2\,, \quad (\lambda=0)\,. \label{eqn:Higuchi} 
\end{equation}
Because of the condition~(\ref{eqn:mT2}), the no-ghost condition (\ref{eqn:noghost-scalar}) is also  independent of the environment.

\vskip.25cm
\noindent{\bf\em Discussion.} This letter describes an extension of the bigravity framework that allows the mass of the graviton to become environmentally dependent by relying on the chameleon mechanism. It offers a novel way to reconcile massive gravity with local tests of General Relativity without invoking the Vainshtein mechanism. This shed a new light on the interplay between local and global tests of gravity, since deviation is expected on cosmological scales at early times. This extension gives more freedom in the parameter space, so that one can expect signatures on cosmological scales while Solar system constraints are satisfied. More generally, this demonstrates the importance of the screening mechanism in drawing the predictions.

In the context of bimetric theories, the stability of the homogeneous and isotropic universe requires (see footnote~\ref{footnote:generalizedHiguchibound}) the generalized Higuchi bound condition~(\ref{eqn:generalHiguchi}) to be satisfied. In the standard bigravity theory on a de Sitter background spacetime, it reduces to the classical Higuchi bound (\ref{eqn:Higuchi}).  In the standard bigravity theory, unfortunately, the bound is violated in the early universe with sufficiently large $H$.  In the chameleonic extension of bigravity proposed in this letter, on the contrary, $m_{\mathrm{T}}$ depends on the environment in such a way that $m_{\mathrm{T}}^2$ scales with the matter energy density,  $m_{\mathrm{T}}^2 \propto \rho$. This significantly broadens the regime of applicability of the bimetric theory. Indeed, for de Sitter backgrounds the ratio $m_{\mathrm{T}}^2/H^2$ does not depend on the vacuum energy density in the Jordan frame (see Eq.~(\ref{eqn:mT2})) and thus the bound (\ref{eqn:noghost-scalar}) is automatically satisfied at all scales once it is satisfied at one scale. For example, if the late-time universe dominated by dark energy is stable then the inflationary universe in the early universe is also stable for the same choice of the parameters of the theory. It is also obvious that the strong coupling scale, $\sim (M_g m_{\mathrm{T}}^2)^{1/3}$, is always higher than the expansion rate $H$.

During the radiation-dominated era, one can find a scaling solution in which $\xi$ and $c$ ($\ne 1$) stay constant and yet each term of (\ref{eqn:Friedmann-eqs1}) scales as $1/a^4$. Actually, one can verify that both these two variables remain approximately constant during the whole evolution of the universe. This scaling solution is an attractor of the system (\ref{eqn:Friedmann-eqs1})-(\ref{kg1}) under a certain condition and leads to the constancy of the ratio $m_{\mathrm{T}}^2/H^2$ again. The cosmological evolution and its stability all the way from the early radiation-dominated epoch to the present epoch with acceleration was shown after submission of the present paper in~\cite{DeFelice:2017gzc}. 

In conclusion, the extension of bigravity models proposed in this letter is very generic and robust. Indeed we have considered only a simple example. One can easily allow for different scalings for the $\beta_i$. Also, the coupling to the field may not be universal and, e.g.\ may be different for the dark matter sector, which would also help to evade local and cosmological constraints simultaneously. It is also possible to generalize the relation between the Einstein frame and the Jordan frame from the simple conformal transformation $\tilde g_{\mu\nu} = A^2(\phi)g_{\mu\nu}$ to more elaborated one such as a disformal transformation of the form $\tilde g_{\mu\nu} = A^2(\phi)g_{\mu\nu}+B(\phi)\partial_{\mu}\phi\partial_{\nu}\phi$. We hope to detail all these issues in our future publication. 

\vskip.25cm
\noindent{\bf\em Acknowledgements.} The authors thank Michele Oliosi and Yota Watanabe for useful comments. SM thanks warm hospitality at IAP, where this work was initiated during his stay. JPU thanks YITP for hospitality. ADF was supported by JSPS KAKENHI Grant Numbers 16K05348, 16H01099. The work of SM was supported by Japan Society for the Promotion of Science (JSPS) Grants-in-Aid for Scientific Research (KAKENHI) No. 24540256, No. 17H02890, No. 17H06359, No. 17H06357, and by World Premier International Research Center Initiative (WPI), MEXT, Japan. The work of JPU is made in the ILP LABEX (under reference ANR-10-LABX-63) was supported by French state funds managed by the ANR within the Investissements d'Avenir programme under reference ANR-11-IDEX-0004-02. 


\end{document}